\documentclass[twoside]{photon2007}
\usepackage[latin1]{inputenc}
\usepackage[dvips]{graphicx,epsfig,color}
\usepackage{wrapfig,rotating}
\usepackage{amssymb,amsmath,array}

\begin{document}

\begin{figure}[htb]
\includegraphics{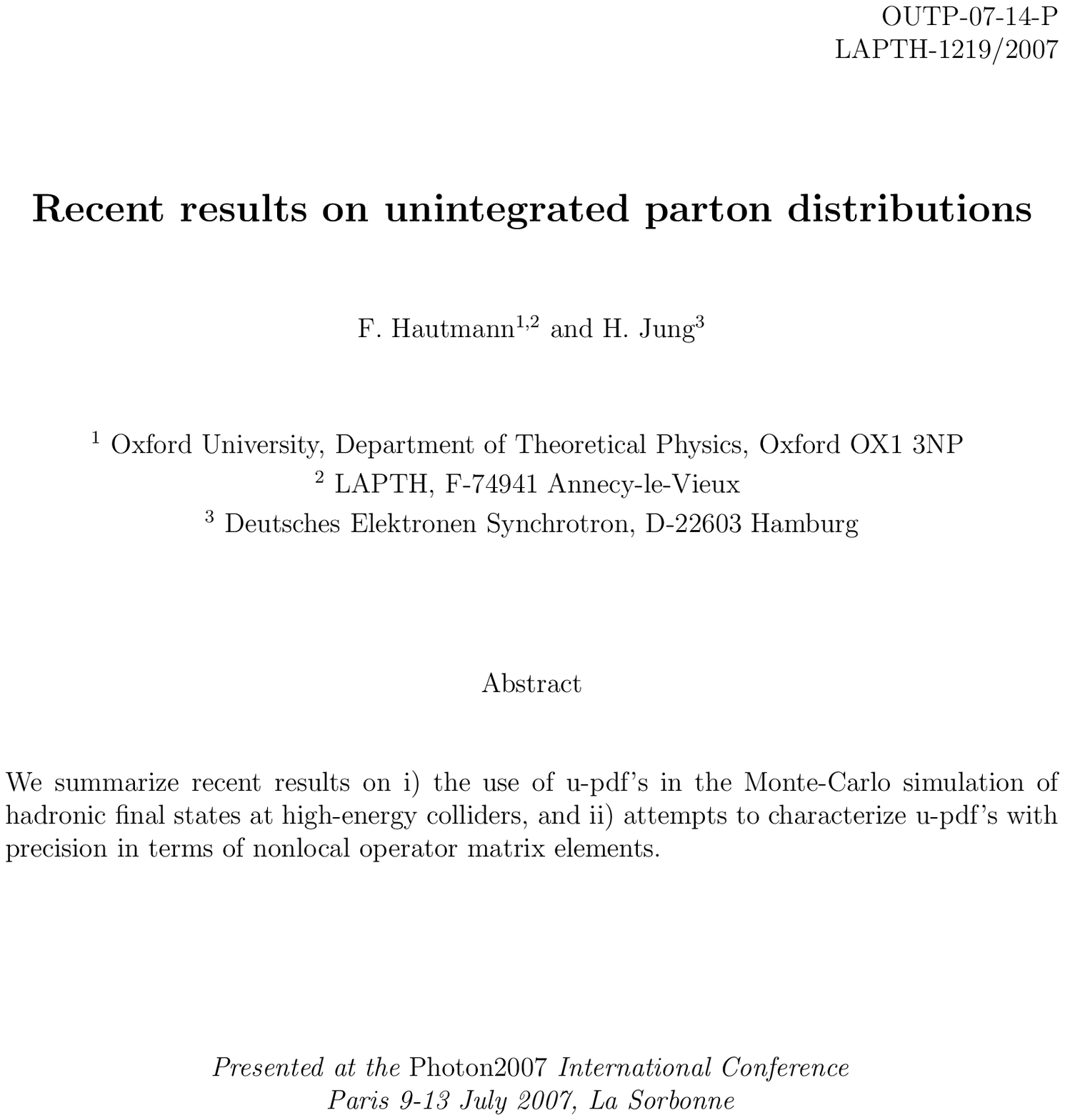}
\end{figure}

\title{
Recent results on unintegrated parton distributions } 
\author{F.~Hautmann$^{1,2}$ and H.~Jung$^3$
\vspace{.3cm}\\
1-   Oxford University, 
    Department of Theoretical Physics, Oxford OX1 3NP
\vspace{.1cm}\\
2-  LAPTH, F-74941 Annecy-le-Vieux   
\vspace{.1cm}\\
3-  Deutsches Elektronen Synchrotron, D-22603 Hamburg
}

\maketitle

\begin{abstract}
We  summarize recent results 
on i) the use of u-pdf's 
 in the Monte-Carlo simulation of 
hadronic final states at high-energy colliders,  
 and ii) attempts to characterize 
u-pdf's  with precision 
in terms of nonlocal operator matrix elements. 
\end{abstract}

\section{Introduction}

At forthcoming high-energy colliders 
 a large amount of events are  
 characterized by  complex final 
states with {\em multiple} hard scales,  possibly 
far apart from each other. 
 QCD methods to address multi-scale 
 hadronic processes involve the use of 
parton distributions unintegrated in both longitudinal and  
transverse components of parton's momentum (u-pdf's). 
 Classic examples are provided by 
Sudakov physics~\cite{ddt,css} and small-x 
physics~\cite{glr,hef90}. 
 U-pdf's also govern  exclusive processes~\cite{brodlep} 
and spin~\cite{mulders}. They are required, in general,  to 
explore detailed features of hadronic final 
states~\cite{heralhcproc}. 
For these reasons, theoretical and phenomenological studies of 
u-pdf's are being actively pursued. 

In the case of small x, u-pdf's can be 
introduced  in a gauge-invariant manner  
 using  high-energy factorization~\cite{hef}. 
This result was used early on both for   
  Monte-Carlo simulations~\cite{marchweb92} of  x~$\to 0$ 
parton showers and for 
  numerical  resummation programs~\cite{ehw} 
for  $\ln$~x corrections to QCD  evolution 
equations. 
For structure function's evolution,    
methods are being 
developed (\cite{ccss07}-\cite{rdball07}  
and references  therein) to   match     
 the k$_\perp$-dependent, small-x dynamics 
with perturbative 
collinear dynamics. 
For the full 
simulation of exclusive components of  hadronic final 
 states, on the other hand,  such matching is   more 
complex (see  discussions in~\cite{heralhcproc} and~\cite{jeppe06})  
and is not yet available. 
This  will be  
critical  for turning present  event generators based on 
u-pdf's~\cite{krauss-bfkl}-\cite{lonn} 
into general-purpose Monte-Carlo tools. 
Importantly,  for x $ \to $ 0 
 u-pdf's may  provide a useful  framework also for  
discussing the approach to the saturation regime:  
see e.g.~\cite{ianmue07,gelvenugo07} for recent studies. 

In the general case,  to characterize 
u-pdf's gauge-invariantly  over the whole 
phase space is  more difficult.  
Open questions on this issue are  the 
    subject of much current 
activity, see  e.g.~\cite{rogers}-\cite{ceccopieri},  
with wide-ranging applications from semi-inclusive processes 
to parton-shower  
algorithms to spin physics. 

In this  report we concentrate on two topics. 
In Sec.~2  we  consider x $\ll$ 1  and  
discuss recent  results  from k$_\perp$ Monte-Carlos 
 on the jet structure of small-x final states. 
 In Sec.~3 we describe  ongoing 
  progress towards 
general operator definitions  for u-pdf's, focusing on 
results for  the treatment  of x $ \to $ 1 endpoint 
 divergences.

\section{Jet final states from unintegrated parton distributions}

There have recently been new calculations 
of hadronic jet final states from Monte-Carlo 
generators based on unintegrated parton distributions. 

\subsection{U-pdf's and shower Monte-Carlo  generators} 

The idea common to the Monte-Carlo event generators based on  
u-pdf's is to use   
 factorization at fixed k$_\perp$~\cite{hef} in order to 
a)~generate the 
hard scattering event, including 
  dependence on the initial 
transverse momentum, and 
b)~couple this to the  evolution of the  
initial state to simulate the   gluon cascade. 
Event generators of this kind include  
{\small SMALLX}~\cite{marchweb92}, 
{\small CASCADE}~\cite{jungcomp}, 
{\small LDCMC}~\cite{sjo}, 
and the newer tools~\cite{krauss-bfkl} and~\cite{golec-mc}. 
Different generators employ different models for the 
evolution of the initial state, such as {\small BFKL}~\cite{bfkl1,bfkl2},  
{\small CCFM}~\cite{mc88,cfm},  {\small LDC}~\cite{lundldc} evolution. 
With suitable constraints on the angular 
ordering of gluon emissions, these can all be set up 
to  give the correct leading 
$\ln x$ behavior. A  number of 
subleading contributions is however also of  importance for 
realistic simulations of final states. 
The general approach of these Monte-Carlos  and 
aspects of the differences among them are reviewed in~\cite{jeppe06}. 
See also~\cite{heralhcproc} for further  references.

Although none of the above generators are nearly 
as developed as standard 
 Monte-Carlos  like {\small PYTHIA}~\cite{pythref} or 
{\small HERWIG}~\cite{herwref}, they have the potential 
advantage of a better treatment  of 
  high-energy logarithmic effects. Implementing 
  these effects  in the shower  
  can be relevant 
 for the simulation of complex  final 
 states at LHC energies, see e.g. multi-jet studies 
in~\cite{mlmhoche}, and 
additional references in~\cite{heralhcproc}.  
 A further potential 
 advantage is that   Monte-Carlos with 
 u-pdf's  likely 
  provide a more natural framework~\cite{gustafstalk}
   to simulate the k$_\perp$  distribution of 
the  soft underlying event~\cite{mcmeet} (minijets, soft 
 hadrons) and,  possibly, multiple interactions 
at very high  energies. 
   
On the other hand, 
present  Monte-Carlos of this kind  do not  
 automatically include 
 contributions from  collinear radiation 
 associated to  x~$\sim 1$. 
They  need to  be corrected for this. 
This is done  partially in present implementations,  but not yet in a 
 systematic fashion. 
A recent example of a study in this direction 
 is in~\cite{krauss-bfkl}, based 
on the procedure  of~\cite{watt,kimber}. 

A related issue concerns the  inclusion of  quark contributions  
 in the initial state. 
In present implementations these are either neglected     
or approximated by lowest-order perturbative 
evolution~\cite{krauss-bfkl,golec-mc,sjo}. 
The   k$_\perp$-dependent 
  kernel  that governs 
 small-x sea-quark evolution  to all orders 
is given in~\cite{ch94}.  Universality properties of this kernel 
are emphasized in~\cite{ccss06,cc05}. The quark kernel  
is not yet implemented in current Monte-Carlos.

Keeping in mind these limitations, 
Monte-Carlo event generators with u-pdf's can be  used at present 
within  the low-x domain   
to probe  the  physical picture of 
space-like x~$\to 0$ parton showers. 
A selection of 
 recent  results from these event generators is given 
 in the next subsections, going 
from  more inclusive  to less inclusive measurements.

\subsection{Inclusive  cross sections}

Available 
k$_\perp$ Monte-Carlos have been used to analyze data 
 for 
DIS structure functions and  for inclusive 
 lepto- and hadro-production of 
jets, heavy flavors and 
prompt photons,  see~\cite{heralhcproc,jeppe06}. 
 These inclusive  analyses serve  to 
test the overall consistency of the  physical picture 
 and to perform first determinations of  
the unintegrated gluon 
distribution.

A consistent semi-quantitative picture is  achieved with 
sensible results for the 
evolved unintegrated gluon 
distribution~\cite{jeppe06,junghgs,watt,kwie,kniehl,szcz07,jungkoti}. 
However,  the 
gluon at low scales and low x 
is  only  poorly constrained 
by this approach~\cite{jungkoti,jungdis04}. 
This is not unexpected, as the  
summation  of higher orders 
implied by the  Monte-Carlo 
should reduce the sensitivity of 
predictions at high scales 
to the form of the input~\cite{angjet}. 

\begin{figure}[htb]
\vspace{60mm}
\includegraphics{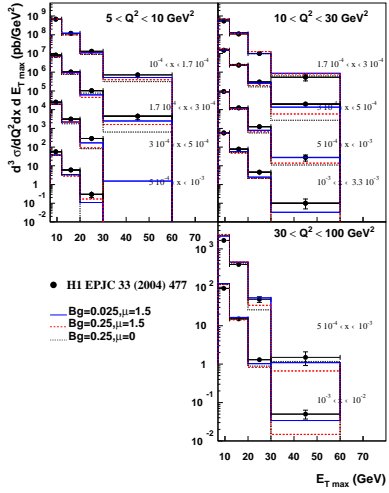}
\caption{Jet $E_T$ distribution at HERA  from 
the k$_\perp$ Monte-Carlo {\small CASCADE} and 
u-pdf fits\protect\cite{junghanss}, compared with H1 
data\protect\cite{h1etjet}.} 
\label{fig:jetET}
\end{figure}

\begin{figure}[htb]
\vspace{45mm}
\includegraphics{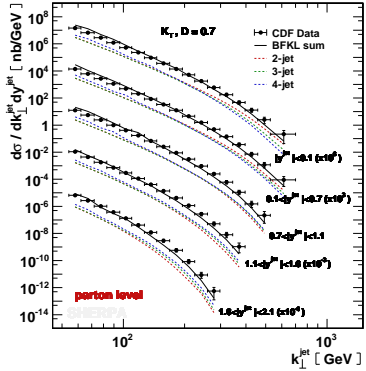}
\caption{Tevatron jet spectra from\protect\cite{krauss-bfkl} 
compared with CDF data\protect\cite{cdfjet07}.} 
\label{fig:jetcdf}
\end{figure}

Examples of 
results 
for jet production in $ep$ and $p {\bar p}$ collisions are 
reported in 
Figs.~\ref{fig:jetET} 
and~\ref{fig:jetcdf}. 
Fig.~\ref{fig:jetET} shows the description 
 of the  jet $E_T$ distribution data 
from H1~\cite{h1etjet}   obtained with 
 {\small CASCADE}  
using the recent u-pdf fits~\cite{junghanss}. 
Fig.~\ref{fig:jetcdf} shows the comparison of CDF 
  jet data~\cite{cdfjet07} with 
the Monte-Carlo~\cite{krauss-bfkl}.  

Current  uncertainties on the unintegrated gluon 
 are  large especially  in  the lowest x region~\cite{jeppe06}. 
Implications of this are   
studied  for heavy-quark structure 
functions at HERA in~\cite{jungkoti} and for jet production 
at RHIC in~\cite{szcz07}. 
Note that first fits of {\em both} the $x$- and 
k$_\perp$-dependence of the unintegrated gluon 
 are performed in~\cite{junghanss}. 

A  potentially important application of these results 
 is to final states containing 
   Higgs bosons at the LHC. 
The  Higgs  matrix element coupled to the unintegrated gluon 
distribution is computed in~\cite{higgs02}. 
 First results from  Monte-Carlo implementation are given  
 in~\cite{junghgs}. Studies of the Higgs  transverse-momentum 
spectrum from  NLO + soft-gluon 
resummation~\cite{vogel,kulesza}  
  indicate that 
 small-x terms, although  
 highly subleading from the viewpoint of the 
 soft hierarchy,  appear to   affect the  spectrum 
  at a level comparable to the current 
theoretical uncertainties estimated as in~\cite{balazs}. 
 Further investigations of these 
effects~\cite{sjo,alekhin13,kwiehiggs,zotovhiggs,szc}   
are warranted. 
An additional,   special application  
is to central exclusive  production, 
recently considered   
for Higgs~\cite{dursusy,durhgs} and 
scalar $\chi$ mesons~\cite{szczter07} in the context of 
u-pdf's. 

\begin{figure}[htb]
\vspace{75mm}
\includegraphics{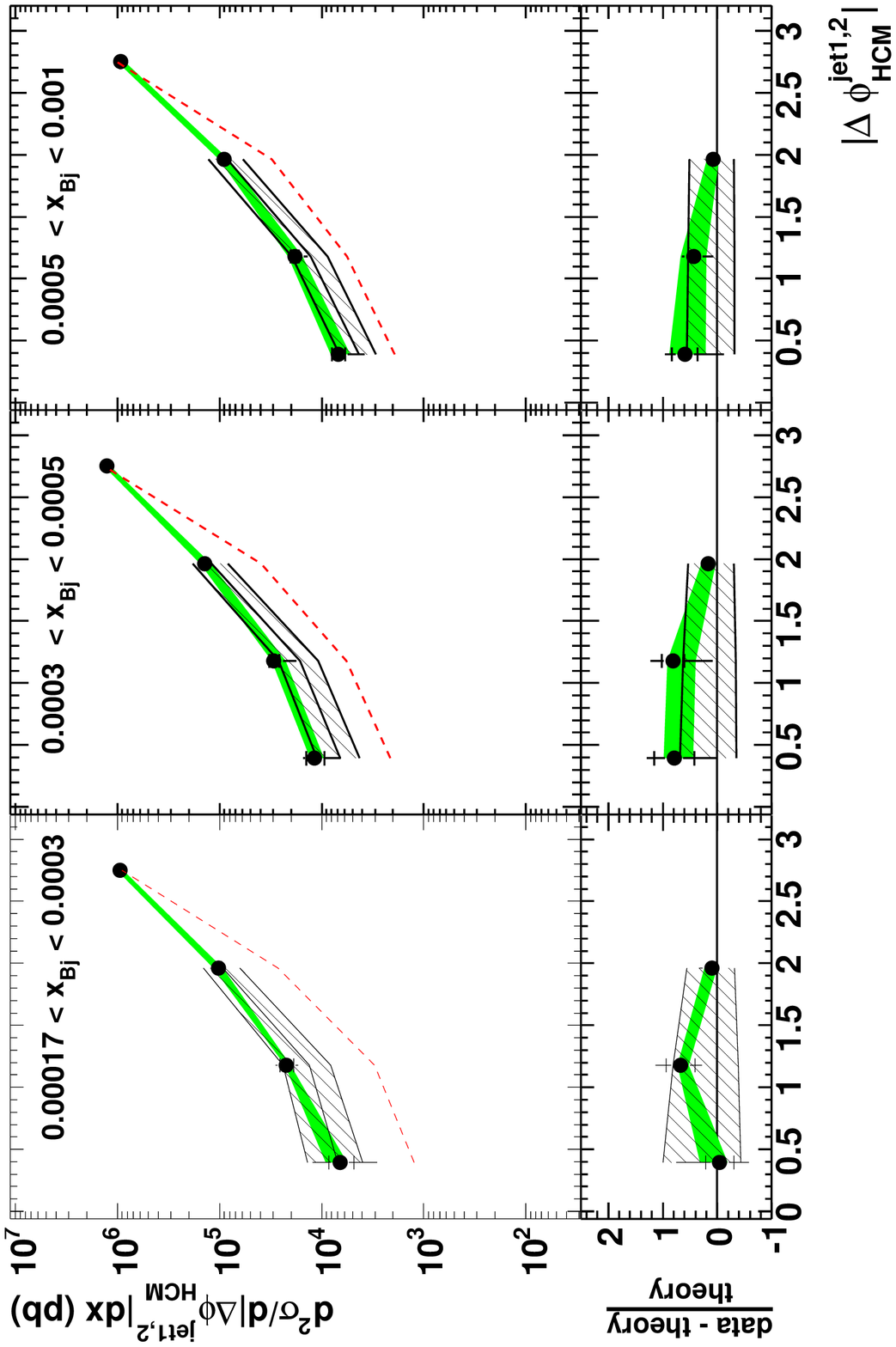}
\includegraphics{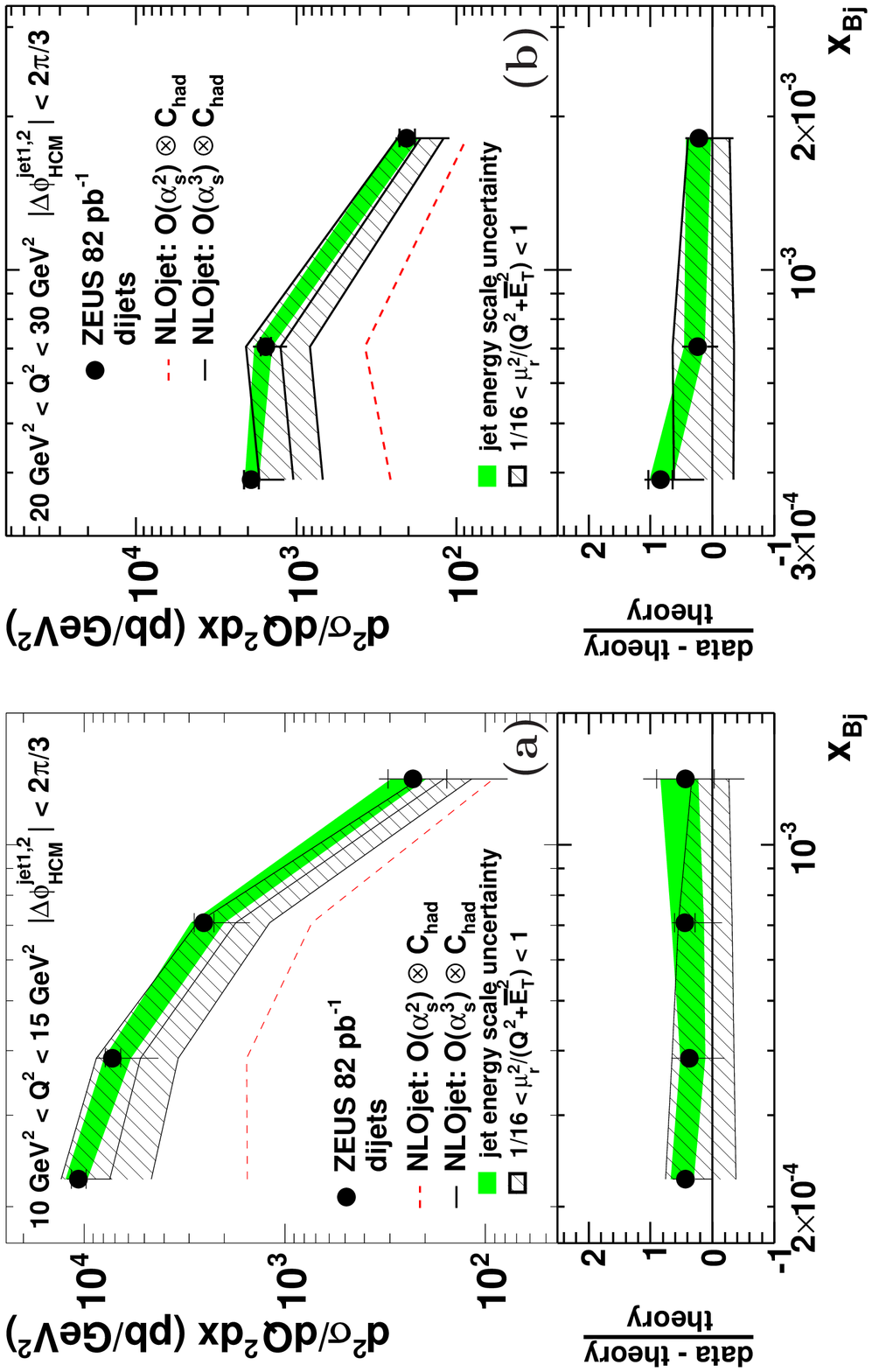}
\caption{(top) Azimuth and 
(bottom) Bjorken-x dependences  
of di-jet distributions 
 measured by Zeus\protect\cite{zeus1931}.} 
\label{fig:phizeus}
\end{figure}

We conclude this 
subsection by observing  that recently the small-x 
definition~\cite{hef} 
of u-pdf's has also been used~\cite{bartels,schwe} 
to treat 
nonleading corrections to jet production at 
large rapidity separations in 
hadron-hadron collisions~\cite{stirvdd}, and 
 Monte-Carlo event generators based on 
 the evolution~\cite{bfkl1,bfkl2} 
are also becoming available for these processes.

\subsection{Multi-jet correlations}

More details of the   parton k$_\perp$ dynamics 
can be probed by examining jet correlations in final states 
containing multiple jets.

Recently the Zeus collaboration has presented measurements of 
two-jet and three-jet distributions in DIS 
associated with low x, 
 $10^{-4} < x < 10^{-2}$~\cite{zeus1931}. 
Zeus' jet definition is such that 
  nonglobal logarithms from 
jet clustering are avoided~\cite{nonglobjet}, as are 
 double logarithms from 
symmetric cuts~\cite{dasguban}. 
Fig.~\ref{fig:phizeus}~\cite{zeus1931} shows  
 two-jet   data versus  the azimuthal 
separation $\Delta \phi$ and  versus Bjorken x,  
compared  with 
next-to-leading-order perturbative results~\cite{nagy}. 
Notice the large variation  
from the 
order-$\alpha^2_s$ to the order-$\alpha^3_s$ result 
with decreasing $x$ and decreasing $\Delta \phi$.

\begin{figure}[htb]
\vspace{75mm}
\includegraphics{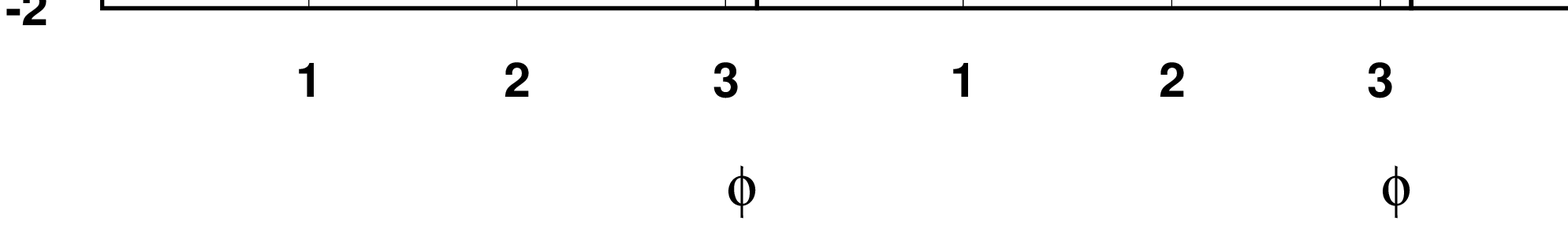}
\includegraphics{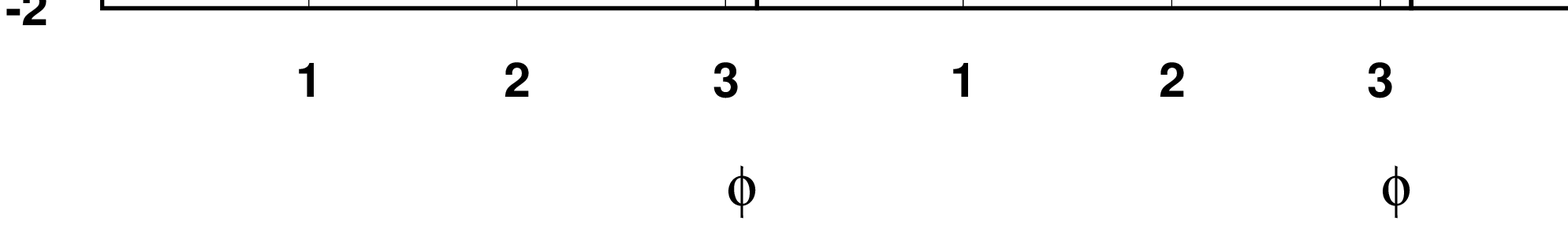}
\caption{Angular jet 
correlations\protect\cite{angjet} 
from {\small CASCADE} and {\small HERWIG}   
compared with Zeus data\protect\cite{zeus1931}: 
(top) di-jet cross section;   
(bottom) three-jet cross section.} 
\label{fig:phipage1}
\end{figure}

\begin{figure}[htb]
\vspace{75mm}
\includegraphics{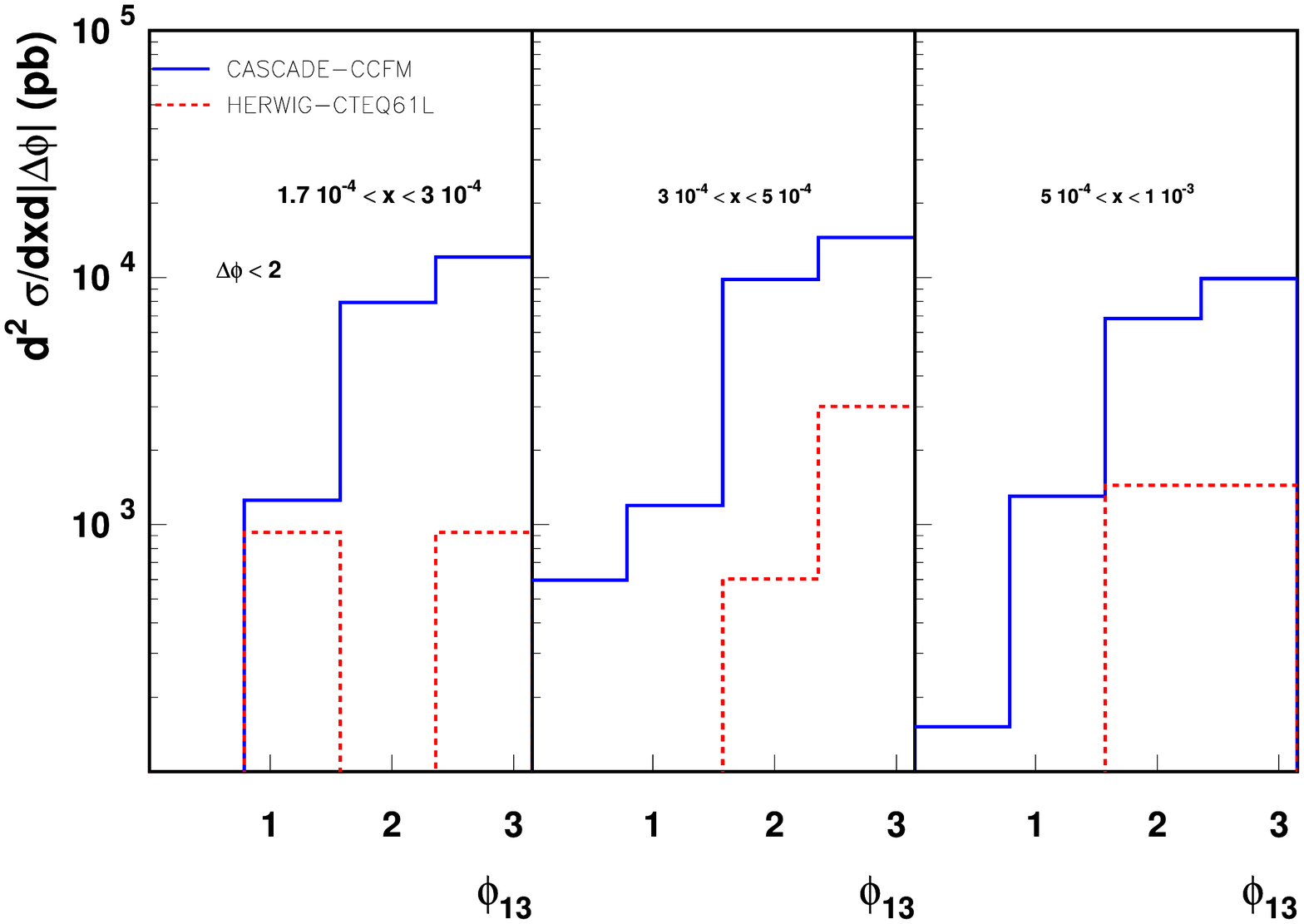}
\includegraphics{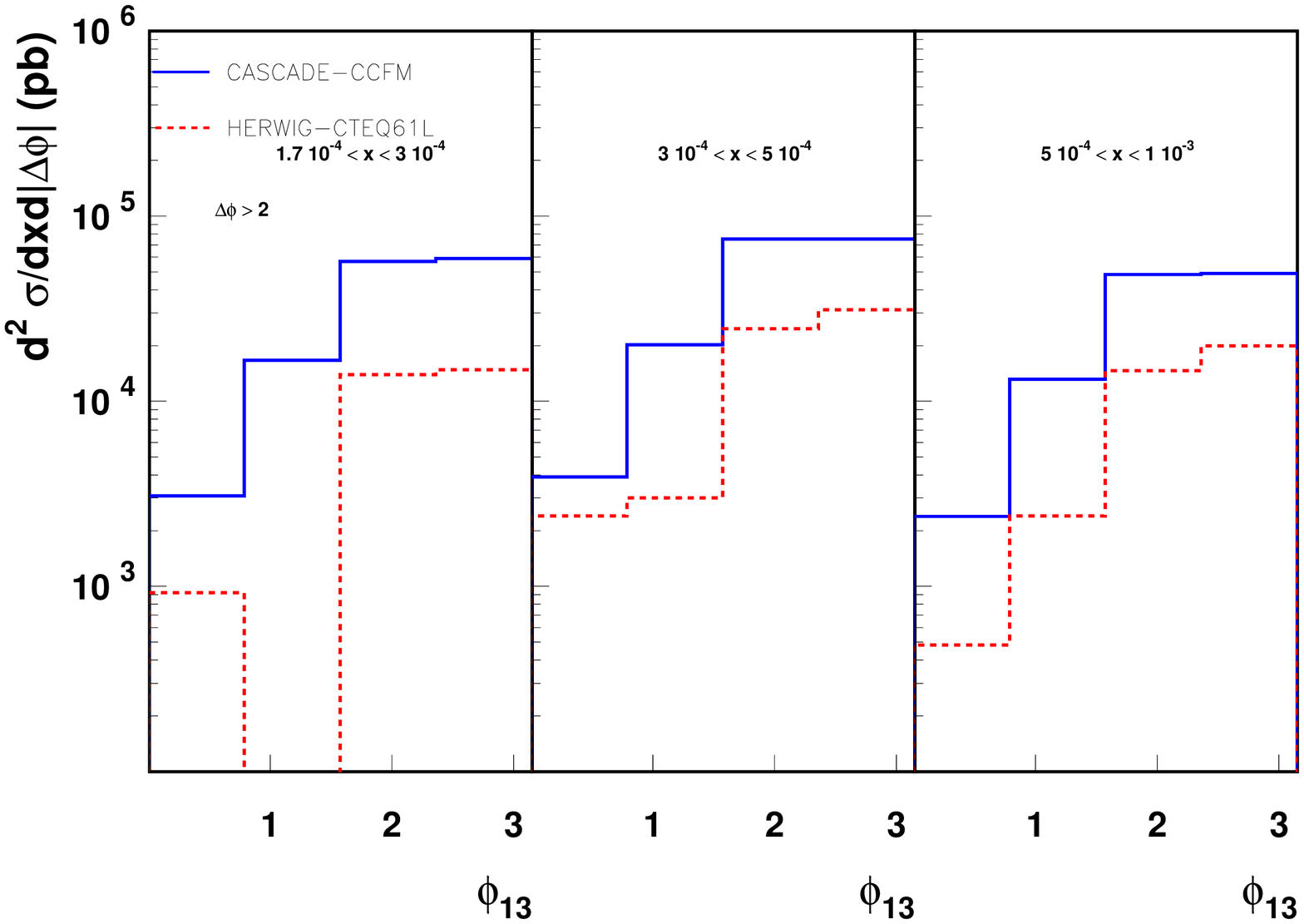}
\caption{Cross section in the azimuthal angle 
between the hardest and the third jet\protect\cite{angjet},   
for (top) small and (bottom) large azimuthal separations 
between the leading jets.} 
\label{fig:thirdjet}
\end{figure}

Ref.~\cite{angjet} analyzes the  multi-jet final states 
in the framework of u-pdf's. In particular, it  
 studies the 
 contribution to potentially large higher-order 
 corrections 
arising from a sizeable k$_\perp$ 
 in the initial state 
when several well-separated hard jets are produced. 
In this kinematics,   
effects from the x~$\ll$~1 parton shower may be 
 enhanced. 
These contributions are  important 
at small $\Delta \phi$, when the jets are not close to  
  back-to-back configurations~\cite{delenda}.  
Fig.~\ref{fig:phipage1} shows results  for 
the distribution in the azimuthal separation 
between the leading jets~\cite{angjet}, compared with the 
measurement~\cite{zeus1931}. 
Fig.~\ref{fig:thirdjet}~\cite{angjet} shows the angular 
distribution for the third jet, in the cases of small 
separations and large separations between the leading jets. 
\noindent The k$_\perp$ Monte-Carlo {\small CASCADE}  
gives a good description of the measurement, and 
gives large 
differences to the collinear-based parton shower   
as implemented in {\small HERWIG}, particularly  
in the region where the azimuthal separations  
 $\Delta \phi$ between the 
leading jets are small.  

\begin{figure}[htb]
\vspace{40mm}
\includegraphics{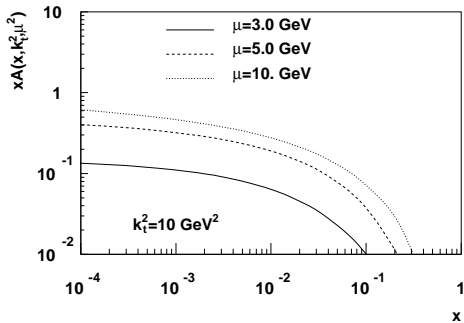}
\caption{The x dependence of the unintegrated gluon 
distribution.} 
\label{fig:ktglu}
\end{figure}

The x dependence of the gluon 
distribution 
used in the above results for the jet cross 
sections is shown 
in Fig.~\ref{fig:ktglu} 
for various scales $\mu$ and fixed k$_\perp$.  

This analysis indicates  that 
 theoretical uncertainties on jet correlations 
 are sizeable at NLO~\cite{zeus1931} and that 
 the k$_\perp$ Monte-Carlo gives an improved description 
 of multi-jets~\cite{angjet} compared to standard parton showers. 
This analysis is for DIS. However, 
despite the much lower energy at HERA, 
note that  
owing to the large phase space available  for jet production
the HERA data may be just as relevant as the Tevatron 
for  understanding the extrapolation to the LHC of initial-state 
radiation effects~\cite{heralhcproc}.

\subsection{Forward region}

Transverse energy flow and  production of jets and 
particles at forward rapidities 
are  important probes of the 
initial-state k$_\perp$~\cite{brw95}. 
Monte-Carlo results for such observables, however,  
strongly depend on the details of the model used for evolution 
of the u-pdf's. 
This model-dependence is much more pronounced for 
forward-region observables than for the jet correlations 
discussed in the previous subsection. 
A  systematic  understanding  of  u-pdf's 
evolution~\cite{collins01} will be  relevant for  more 
precise interpretations of  forward 
measurements. 
See~\cite{ceccopieri} for recent work 
on evolution equations, including  discussion 
of  target fragmentation. 
We refer the reader to~\cite{heralhcproc,jeppe06}, 
and references therein, for 
discussion of current forward-region results. 

\section{Towards precise characterizations of u-pdf's}

 The possibility to turn   the event  generators  discussed in 
  Sec.~2  
into general tools to describe    hadronic final states over 
the whole phase 
space   depends  on theoretical progress 
about  unintegrated parton distributions. 
This will involve  
precise characterizations of the distributions, factorization and 
evolution equations. In what follows we 
discuss recent progress 
in the understanding of   
operator  matrix elements for u-pdf's and their 
 lightcone limits.

\subsection{Gauge invariant matrix elements}

The relevance of consistent operator definitions for 
parton k$_\perp$ distributions  was emphasized 
long ago, see e.g.~\cite{brodlep,collsud}. 
Ensuring gauge invariance at k$_\perp \neq 0$, however, 
 is nontrivial. 
 To this end  
 the  approach commonly used 
   is to  generalize     
 the coordinate-space  matrix elements  that serve to 
 identify   ordinary pdf's 
to the case 
of field operators  at non-lightcone distances. 
E.g., for quarks one has~\cite{mulders,beli,jccrev03}  
\begin{equation}
\label{coomatrel}
  {\widetilde f} ( y  ) =
  \langle P |  {\overline \psi} (y  )
  V_y^\dagger ( n ) \gamma^+ V_0 ( n )
 \psi ( 0  ) |
  P \rangle  \hspace*{0.3 cm}  .   
\end{equation}
Here  $\psi$ are  the quark fields evaluated at distance 
$y = ( 0 , y^- , y_\perp  )$,   $y_\perp$ is in general nonzero, 
 and   $V$ are  eikonal-line operators in direction $n$ required 
 to make the matrix element gauge-invariant. 
 The unintegrated quark distribution is obtained from the double 
 Fourier transform in  $y^-$ and $y_\perp$ of ${\widetilde f}$. 
 An extra  gauge link at infinity~\cite{beli} 
is to be taken into account  in the case of physical  gauge.

There are subtleties, however,  to using 
Eq.~(\ref{coomatrel}) beyond tree level.  As realized 
early on~\cite{collsud}, parton distributions at 
 fixed k$_\perp$ are  no longer protected 
  by the KLN mechanism~\cite{kln} 
  against uncancelled lightcone divergences near 
  the x $=$ 1 endpoint~\cite{jccrev03,brodsky01}. 
It is only after supplying the above matrix element 
with a regularization prescription that the distribution  is 
well-defined. 

Importantly, similar to what 
observed  in~\cite{jccfh00}  for the case of the Sudakov 
form factor, the choice of a particular 
regularization method for the lightcone divergences   also 
affects  the distributions  
integrated over k$_\perp$ 
and the ultraviolet subtractions.
A further  set of  questions 
 being studied~\cite{fhfeb07,watt,jcczu,cch93} are in fact   
  associated with   the  coefficient functions 
  that govern the expansion 
  of  unintegrated distributions in terms of ordinary ones, and the 
  relation of the integral of u-pdf's with ordinary 
  distributions.

\begin{figure}[htb]
\vspace{25mm}
\includegraphics{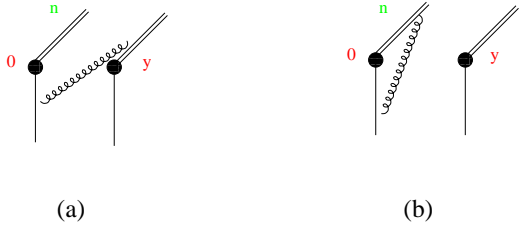}
\caption{One-loop contributions to lightcone divergences 
in the matrix element (\ref{coomatrel}). } 
\label{fig:wloop1}
\end{figure}

The above issues can be   analyzed by explicit calculation 
at one 
loop~\cite{fhfeb07,fhdistalk}.  
 Expansion in powers of $y^2$ of the 
 result for the coordinate-space matrix 
element  at this order  (Fig.~\ref{fig:wloop1})
yields 
\begin{eqnarray}
\label{EbmEaexpand}
&& {\widetilde f}_{1} (y)  =   
{{ \alpha_s C_F     } \over { \pi  } } 
\ p^+
\int_0^1 dv  \ { v \over { 1 - v }} \ 
\nonumber\\
&& \times 
\left\{ 
\left[ e^{ i p \cdot y v}  - e^{ i p \cdot y } \right] 
\ \Gamma ( 2 - { d \over 2} ) \ 
( { {4 \pi \mu^2} \over \rho^2} )^{2-d/2}
\right. 
\nonumber\\
 && + 
e^{ i p \cdot y v} \ \pi^{2-d/2} \ 
\Gamma (  { d \over 2} - 2 ) \ 
(- y^2  \mu^2)^{2 - d/2}
\nonumber\\
 && + \left.
 \cdots 
\right\} \hspace*{0.2 cm}  , 
\end{eqnarray}
where $\mu$ is the  dimensional-regularization scale 
  and $\rho$ is an infrared  mass regulator.  
The lightcone singularity $v \to 1$,  
corresponding  to the exclusive  boundary $x = 1$,   
  cancels for ordinary pdf's  
 (first term in the right hand side of Eq.~(\ref{EbmEaexpand})) 
but it    
 is present, even at $d \neq 4$ and finite 
$\rho$,   in subsequent terms. 
This implies that, using 
the matrix element (\ref{coomatrel}), the $1/ (1-x)$ factors from 
 real emission probabilities do not combine with 
 virtual corrections to give  $1/ (1-x)_+$ 
 distributions, but leave uncancelled divergences at fixed 
 k$_\perp$.

Different methods have been employed in the literature to 
provide u-pdf's with consistent regularization 
for the endpoint region. We 
consider these methods briefly in the next subsection.

\subsection{Cut-off vs. subtractive method} 

A possible approach to treat the endpoint  is to 
implement a cut-off by taking the eikonal line $n$ to be 
non-lightlike, 
as in the early works~\cite{collsud} and~\cite{korch89}.  
The cut-off in x at fixed k$_\perp$ is of order 
$1 - x \geq k_\perp / \sqrt{4 \eta}$, where 
$\eta = (p \cdot n)^2 / n^2$ and $p$ is the incoming momentum. 
Evolution 
equations in the cut-off parameter 
$\eta$ are investigated 
in~\cite{ji06,collsud,korchangle,jiyuan}.  
A thorough analysis of factorization at leading order has 
recently been given in~\cite{rogers} based on this approach. 

A potential drawback of this approach 
is that   cut-off regularization is not very  
well-suited for applications  beyond the leading order. 
Also, as the two lightcone limits $y^2 \to 0$ 
and $n^2 \to 0$ do not commute, the integral over 
 k$_\perp$ of the distribution   has a finite 
 dependence on  the regularization 
parameter $\eta$, 
\begin{eqnarray}
\label{neqrelat}
\int d k_\perp \ f ( x, k_\perp, \mu, \eta )
  &=& F (x,  \mu, \eta )
\\  
  &\neq&   {\rm ordinary}  \hspace*{0.2 cm} {\rm pdf} \hspace*{0.2 cm} , 
\nonumber
\end{eqnarray}
which makes the 
  relation with the   standard 
operator product expansion  not so transparent. 

An  alternative route 
is based on the 
subtractive  method~\cite{jccfh00,jccfh01}. In 
this case    the direction 
$n$ is kept to be lightlike  but the divergences 
are canceled by multiplicative, gauge-invariant counterterms 
given by   vacuum expectation  
values of eikonal operators. 
The counterterms contain in general 
  both lightlike and 
non-lightlike eikonals. 
For this reason 
studies of u-pdf's in this framework  
introduce  auxiliary eikonals  in direction 
$u = ( u^+, u^-, 0_\perp)$ with  
$ u^+$ and  $u^-$  nonzero~\cite{fhfeb07,jccrev03}. 
The  counterterms have  compact  all-order 
expressions in coordinate space.   
At one loop they  provide 
the regularization of the 
 x $\to$ 1 endpoint 
through a k$_\perp \neq 0$ 
extension 
of the plus-distribution regularization, whose   
 specific form  is determined  in~\cite{fhfeb07}.

The subtractive method is 
  more systematic  than the cut-off, and 
likely  more 
suitable for using unintegrated parton distributions 
at subleading-log level.  
In particular, 
the specific choice of the  counterterms~\cite{fhfeb07} 
is such that the dependence on the 
 non-lightlike eikonal $u$ cancels in the matrix element at 
 $y_\perp = 0$, corresponding to the ordinary pdf.   
The subtractive method may  be helpful 
 for global NLO analyses incorporating 
Sudakov resummation~\cite{cpyuan,cpyuan1}, 
  and construction of 
parton-shower algorithms beyond leading order~\cite{jcczu,bauermc}. 
See~\cite{manohstew,leesterm,idimeh} for studies of 
subtractive methods similar to those of~\cite{jccfh00,jccfh01} 
in relation to  effective-theory techniques.

\subsection{Further issues in higher order} 

\begin{figure}[htb]
\vspace{28mm}
\includegraphics{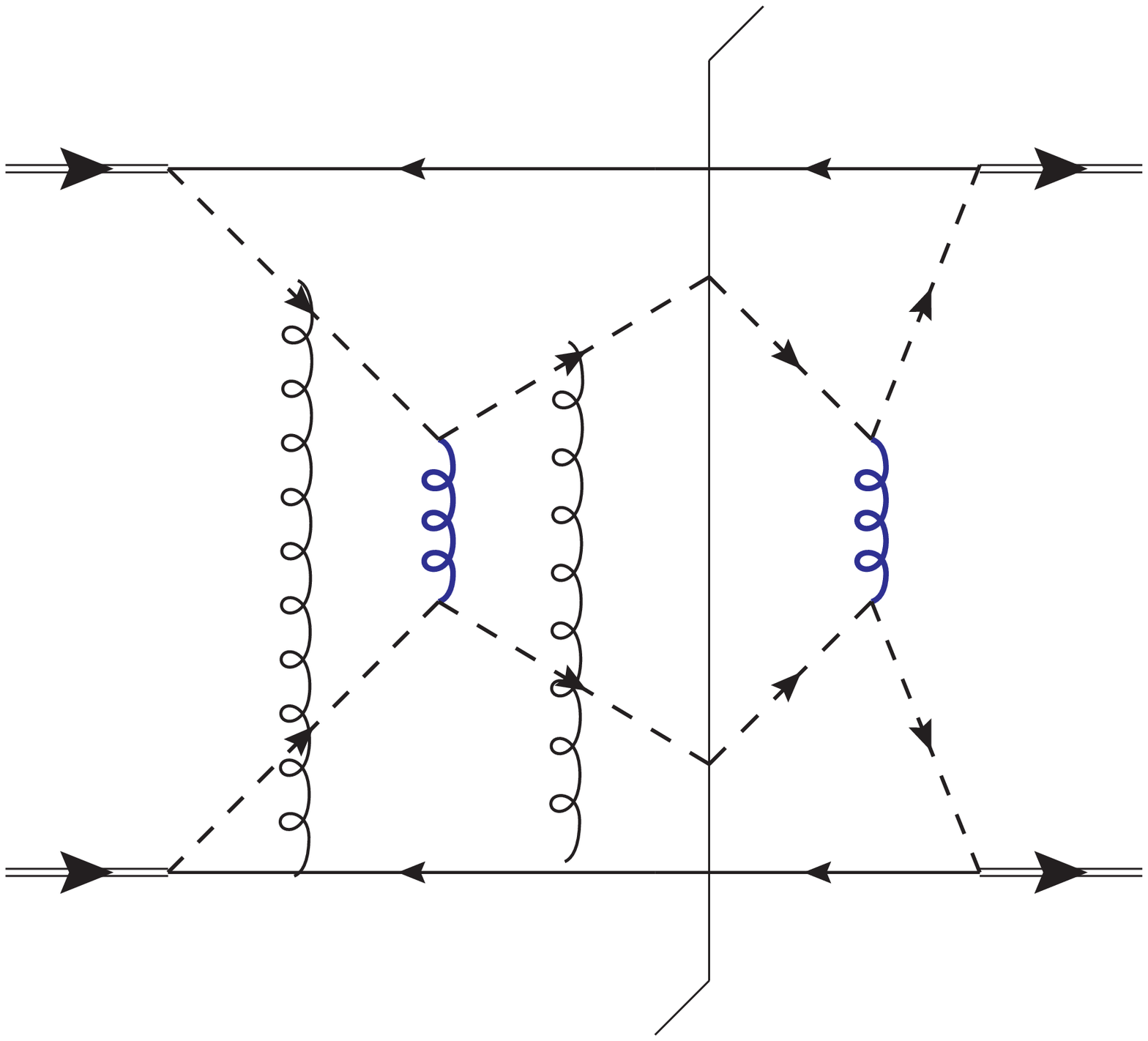}
\caption{Soft gluon exchange with spectator partons.} 
\label{fig:spect}
\end{figure}

Further issues arise 
in higher order. Non-universality of k$_\perp$-dependent parton 
distributions~\cite{pijlman} has  recently been  
studied~\cite{bomhmuld07}-\cite{vogel0708} 
in the hadroproduction of back-to-back high-$p_t$ particles.  
Potential factorization-breaking effects from soft gluons coupling 
initial and final states are shown in 
Fig.~\ref{fig:spect}~\cite{collins0708}. These would enter at a 
high order of perturbation theory (N$^3$LO correction to dihadron 
production).   
Although the  role of these corrections 
 is yet to be fully established, it is  interesting 
that    Coulomb/radiative mixing terms 
also appear to break 
color coherence~\cite{manch}  
in high-order contributions to 
 cross sections for dijets with a gap in rapidity. 
A satisfactory understanding of factorization should 
 likely require a full clarification of these issues. 

\section{Conclusion}

Monte-Carlo event generators 
are being developed 
based on  gauge-invariant definition of 
u-pdf's at small x (Sec.~2.1). 
Compared to standard parton shower approaches, 
these Monte-Carlos have the advantage 
of   
taking into account QCD  initial-state  radiation 
processes that depend on the large-k$_\perp$ 
tail of partonic distributions and 
matrix elements. 

Such processes are likely to be relevant 
to the simulation of complex final states 
characterized by  multiple hard scales. 
An example is the DIS multi-jet correlations 
discussed in Sec.~2.3.  But 
  we expect  analogous considerations to 
   apply 
for multi-jet final states in hadron-hadron 
collisions at LHC energies. 

 k$_\perp$ Monte-Carlos are currently 
 being tuned and validated  
 using   HERA and Tevatron data, as discussed in 
 Secs.~2.2 and~2.3. 
 Determinations 
 of the unintegrated  gluon density are performed.

It will be important to 
extend the notion of u-pdf's beyond x~$\ll$~1 
(Sec.~3) 
 to turn the above event generators 
into general-purpose tools. 

  This involves  a   challenging theoretical 
program, raising  new questions  on operator 
matrix elements (Sec.~3.1),
factorization  and  lightcone 
divergences (Sec.~3.2), and possibly new 
Coulombic effects  (Sec.~3.3).

\begin{footnotesize}

\end{footnotesize}

\end{document}